\documentclass[aps,prb,floatfix,twocolumn,showpacs]{revtex4-1}

\usepackage{graphicx}
\usepackage{color}
\usepackage{amsthm}
\usepackage{stmaryrd}

\begin{document}
\draft
\title{Topological spin Hall and spin Nernst effects in a bilayer graphene}
\author{A. Dyrda\l$^{1}$, J.~Barna\'s$^{1,2}$}

\address{$^1$Department of Physics, Adam Mickiewicz University,
ul. Umultowska 85, 61-614 Pozna\'n \\
$^2$  Institute of Molecular Physics, Polish Academy of Sciences,
ul. M. Smoluchowskiego 17, 60-179 Pozna\'n, Poland}
\date{\today }

\begin{abstract}
We consider intrinsic contributions to the spin Hall and spin
Nernst effects in a bilayer graphene. The relevant electronic
spectrum is obtained from the tight binding Hamiltonian, which
also includes the intrinsic spin-orbit interaction. The
corresponding spin Hall and spin Nernst conductivities are
compared with those obtained from effective Hamiltonians
appropriate for states in the vicinity of the Fermi level of a
neutral bilayer graphene. Both conductivities are determined
within the linear response theory and Green function formalism.
The influence of an external voltage between the two atomic sheets
is also included. We found transition from the topological spin
Hall insulator phase at low voltages to conventional insulator
phase at larger voltages.
\end{abstract}

\maketitle

\section{Introduction} 

Four decades ago Dyakonov and Perel showed that a system with
strong spin-orbit interaction should reveal transverse spin
current and spin accumulation in the presence of external
longitudinal electric field~\cite{dyakonov, dyakonovlett} --  even
if the system is nonmagnetic. This effect, known now as the spin
Hall effect (SHE)~\cite{hirsch}, was studied extensively in the
last few years~\cite{hirsch, murakami03, sinova, kato, kimura,
brune, engel}, and is still of current interest -- mainly because
it offers a new possibility of spin manipulation with electric
field only. The possibility of pure electrical manipulation of
spin degrees of freedom is interesting not only from fundamental
reasons, but also from the point of view of possible applications
in future spintronics devices and  information processing
technologies~\cite{wolf, zutic, sih, awschalom2009}.

The crucial interaction responsible for SHE, i.e. the spin-orbit
coupling, may be either of intrinsic (internal) or extrinsic
origin. The corresponding extrinsic SHE is associated with
mechanisms of spin-orbit scattering on impurities and other
defects (skew scattering and/or side jump), while the intrinsic
SHE is a consequence of a nontrivial trajectory of charge carriers
in the momentum space due to the spin-orbit contribution of a
perfect crystal lattice to the corresponding band structure. The
intrinsic SHE may be described in terms of the Berry phase
formalism~\cite{berry,sundaram} and therefore it is also referred
to as the topological SHE.

It is well known that various spin effects, like for instance spin
current and spin accumulation, may be generated not only by
external electric field, but also due to a temperature gradient.
Indeed, there is a great interest currently in spin related
thermoelectric effects. One of such phenomena is the spin Seebeck
effect, where longitudinal spin current and spin voltage are
generated by a temperature gradient~\cite{uchida}. Of particular
interest, however,  are spin thermoelectric effects in systems
with spin-orbit interaction, where a temperature gradient gives
rise to transverse spin accumulation and/or spin currents. Thus,
the temperature gradient in such systems may lead to anomalous (in
case of ferromagnetic systems) and spin  Nernst
effects~\cite{xiao2007, czhang2009, jaworski}, which correspond to
the anomalous and spin Hall effects induced by longitudinal
electric field. Similarity of the spin Hall and spin Nernst
effects in nonmagnetic systems with spin-orbit interaction is
presented in Fig.1, which clearly shows that the SHE is generated
by external electric field, while the spin Nernst effect (SNE) is
a similar effect generated by a temperature gradient instead of
electric field (gradient of electrostatic potential).

\begin{figure}[t]
  \includegraphics[width=0.9\columnwidth]{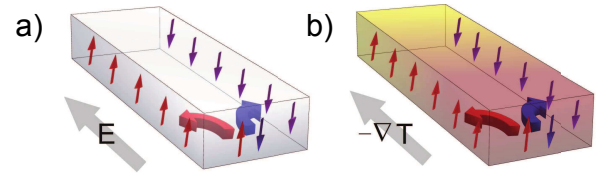}
  \caption{(color on-line) Schematic presentation of
  transverse spin accumulation induced by  (a) electric field
  (spin Hall effect) and (b) temperature gradient (spin Nernst effect).  }
\end{figure}

In this paper we consider the topological contribution to the spin
Hall and spin Nernst effects in a bilayer graphene. Graphene is a
two-dimensional crystal of carbon atoms. A monolayer graphene in a
free standing form was obtained few years ago and owing to its
unusual and peculiar properties quickly became one of the most
extensively studied materials~\cite{Geim2007, katsnelson, castro}.
After the pioneering paper by Kane and Mele~\cite{kane}, spin Hall
effect in a monolayer graphene was studied in many papers and in
various physical situations. Recently, bilayer graphene is
extensively studied as a more appropriate for applications than a
single-layer one. Moreover, it has been shown that spin-orbit
coupling in a bilayer can be enhanced in comparison to that in a
single-layer graphene~\cite{guinea}. This motivated us to consider
topological contributions to the SHE and SNE in a bilayer
graphene.

Both SHE and SNE in (nonmagnetic) graphene are generated by
spin-orbit interaction. In general, one can distinguish two
different forms of the spin-orbit interaction having the crystal
lattice periodicity and contributing to the relevant band
structure -- internal and Rashba spin-orbit interactions. The
latter interaction is due to a substrate and can be controlled by
an external gate voltage. In this paper we consider the
contributions to SHE and SNE due to the intrinsic spin-orbit
interaction only. It is known that this interaction opens an
energy gap at the Dirac points. It was also shown that the energy
gap can be tuned externally by applying a voltage bias between the
layers~\cite{castro2007, zhangtangwang}. The gate voltage
dependence of the spin Hall conductivity, leading to phase
transition between the spin Hall insulator and classical insulator
will also be considered in this paper.

The description of transport properties of graphene is usually
based on some effective Hamiltonian, which properly describes
electronic spectrum near the Fermi level of a neutral system.
However, it is well known that the topological contribution to the
spin Hall effect includes contributions from electronic states far
below the Fermi level, and therefore a more accurate electronic
spectrum is required to describe the effect properly. Accordingly,
in this paper we calculate the spin Hall and spin Nernst
conductivities from a more realistic electronic spectrum based on
a tight binding Hamiltonian, and compare them with the
corresponding conductivities obtained on the basis of effective
Hamiltonians. The results presented in this paper reveal, however,
a very good agreement between the conductivities derived from the
effective and tight binding Hamiltonians.

The paper is organized as follows. In section 2 we describe
briefly the electronic states of a bilayer graphene within the
tight binding Hamiltonian and also in terms of an effective
Hamiltonian, which is sufficient  for states in the vicinity of
the Fermi level of a neutral graphene. In both cases the intrinsic
spin-orbit interaction and the effect of a normal bias voltage are
taken into account. In section 3 we calculate the spin Hall and
spin Nernst conductivities for the tight binding Hamiltonian and
compare them with those obtained from the effective Hamiltonian.
In the latter case we derive analytical formulas for the spin Hall
and spin Nernst conductivities. We also discuss the role of normal
bias (in the framework of the effective model). Description based
on a reduced low-energy effective Hamiltonian is presented in
section 4. Summary and final conclusions are given in section 5.

\section{Electronic spectrum of the bilayer graphene} 

A single-layer graphene is a monolayer of carbon atoms arranged in
a two-dimensional honeycomb lattice which can be also considered
as being composed of two nonequivalent triangular sublattices. In
the absence of spin-orbit interaction the Fermi surface of a
neutral single-layer graphene consists of two nonequivalent $K$
and $K^\prime$ points of the Brillouin zone, at which the valence
and conduction bands touch each other. The corresponding
electronic spectrum can be described by a tight binding
Hamiltonian with nearest and next-nearest neighbor hopping terms.
The low energy electron states near the points $K$ and $K^\prime$
can be well approximated by a conical energy spectrum (linear
dispersion relations). As a result, charge carriers in the
vicinity of the points $K$ and $K^\prime$ are described
effectively by the relativistic Dirac equation.\cite{Geim2007,
katsnelson} Intrinsic spin-orbit interaction opens then an energy
gap at the Dirac points. \cite{kane}. The tight binding and
effective Hamiltonians for a bilayer graphene are more complex, as
described below.

\subsection{Tight binding model}

The bilayer graphene in the Bernal stacking (see {\it eg}
[\onlinecite{castro}]) is described by the following tight binding
Hamiltonian:
\begin{eqnarray}
\mathcal{H} = \int d^{2}{\bf k}\, \psi({\bf k})^{\dag} \left(
\begin{array}{cc}
    H & \Gamma \\
    \Gamma^{\dag} & H
\end{array}
 \right) \psi({\bf k}),
\end{eqnarray}
where
\begin{eqnarray}
H =
\left(\begin{array}{cc}
    h_{so} S_{z} + V S_{0} & h_{0}S_{0} \\
    h^{*}_{0}S_{0} & - h_{so} S_{z} - V S_{0}
\end{array}\right),
\end{eqnarray}
\begin{eqnarray}
\Gamma = \left( \begin{array}{cc}
0 & \gamma_{1}\, S_{0}\\
0 & 0 \\
\end{array}\right).
\end{eqnarray}
The matrix elements $h_{0}$ and $h_{so}$ are defined as follows:
$h_{0} = -t [e^{i 2 b k_{y} /3} + 2 \cos{(\frac{1}{2} a k_{x})}
\,e^{- i b k_{y} /3} ]$ and $h_{so} = -2 t^\prime [ \sin{( a
k_{x})} - 2 \sin{(a k_{x}/2)\,\cos{(b k_{y})}}]$, where $t$ is the
hopping integral between the nearest neighbors in the atomic
sheets, $t^\prime$ is the next-nearest neighbor spin-orbit hopping
amplitude, while $b = a \sqrt{3}/2$ with $a$ being the lattice
parameter.  Furthermore, $V$ is the voltage between the two atomic
sheets of the bilayer (measured in energy units), $S_\alpha$
denote the unit ($\alpha =0$) and Pauli ($\alpha =x,y,z$) matrices
in the spin space, while $\gamma_1$ describes coupling between the
two atomic layers.

\begin{figure}[t]
  \includegraphics[width=0.75\columnwidth]{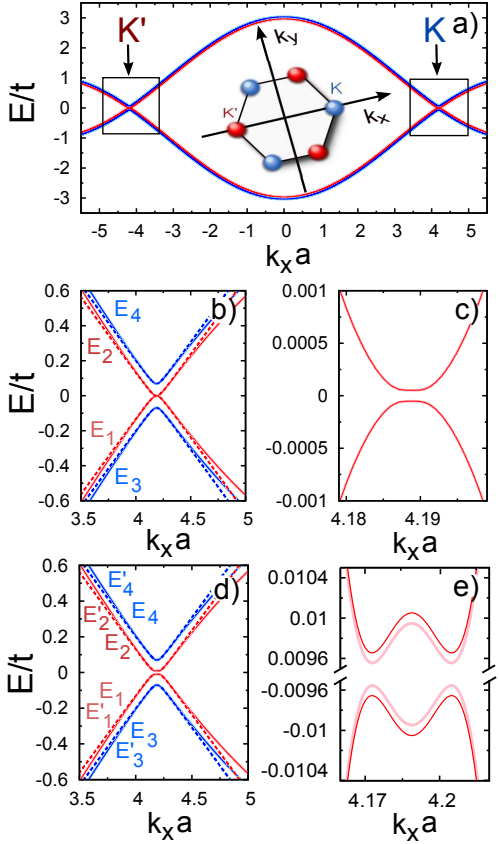}\\
  \caption{Electronic spectrum of the bilayer graphene in the
tight binding model for $k_{y} = 0$, $t=2.9$ eV, $t'/t = 10^{-5}$,
$\gamma_{1}/t = 0.069$, and $V=0$ (a). The tight binding energy
spectrum in the vicinity of the K point (solid lines) and the
corresponding energy spectrum obtained from the effective model
(dotted lines) are compared in (b) for $V=0$ and in (d) for $V/t =
0.01$. Parameters of the effective model, corresponding to the
tight binding one, are: $\Delta_{so}=0.15$ meV and $v=3.516 \times
10^{-10}$ eVm. The electron states very close to the energy gap
are shown in (c) for $V=0$ and in (e) for $V/t = 0.01$ (here the
results from the tight binding and effective Hamiltonians
overlap).}
\end{figure}

The corresponding energy
eigenvalues for $V = 0$ have then the following form:
\begin{eqnarray}
E_{1,2} = \mp \left[h_{so}^{2} + \frac{1}{2}\left(\gamma_{1}^{2} +
2 h_{0}^{2} - \gamma_{1}\sqrt{\gamma_{1}^{2} + 4
h_{0}^{2}}\right)\right]^{1/2}
\end{eqnarray}
\begin{eqnarray}
E_{3,4} = \mp \left[h_{so}^{2} + \frac{1}{2}\left(\gamma_{1}^{2} + 2 h_{0}^{2} + \gamma_{1}\sqrt{\gamma_{1}^{2} + 4 h_{0}^{2}}\right)\right]^{1/2}
\end{eqnarray}
This spectrum is shown in Fig.2(a). States near the point $K$ are
shown by the solid lines in parts (b) and (c). The part (c)
reveals a small energy gap created at the Dirac points by the
spin-orbit interaction.

When $V \neq 0$, the inversion symmetry is broken (layers are no
longer equivalent) and the degeneracy is lifted. The corresponding
eigenvalues acquire then the form
\begin{eqnarray}
E_{1,2} = \mp\left[h_{0}^{2} + h_{so}^{2} + V^{2} +
\frac{\gamma_{1}^{2}}{2}\right. \hspace{3cm}\nonumber\\ -
\left.\frac{1}{2} [(\gamma_{1}^{2} - 4 V h_{so})^{2} + 4
(\gamma_{1}^{2} + 4 V^{2}) h_{0}^{2}]^{1/2}\right]^{1/2},
\end{eqnarray}
\begin{eqnarray}
E_{1^\prime,2^\prime} = \mp\left[h_{0}^{2} + h_{so}^{2} + V^{2} +
\frac{\gamma_{1}^{2}}{2}\right. \hspace{3cm}\nonumber\\ -
\left.\frac{1}{2} [(\gamma_{1}^{2} + 4 V h_{so})^{2} + 4
(\gamma_{1}^{2} + 4 V^{2}) h_{0}^{2}]^{1/2}\right]^{1/2},
\end{eqnarray}
\begin{eqnarray}
E_{3,4} = \mp\left[h_{0}^{2} + h_{so}^{2} + V^{2} +
\frac{\gamma_{1}^{2}}{2}\right. \hspace{3cm}\nonumber\\ +
\left.\frac{1}{2} [(\gamma_{1}^{2} - 4 V h_{so})^{2} + 4
(\gamma_{1}^{2} + 4 V^{2}) h_{0}^{2}]^{1/2}\right]^{1/2},
\end{eqnarray}
\begin{eqnarray}
E_{3^\prime,4^\prime} = \mp\left[h_{0}^{2} + h_{so}^{2} + V^{2} +
\frac{\gamma_{1}^{2}}{2}\right. \hspace{3cm}\nonumber\\ +
\left.\frac{1}{2} [(\gamma_{1}^{2} + 4 V h_{so})^{2} + 4
(\gamma_{1}^{2} + 4 V^{2}) h_{0}^{2}]^{1/2}\right]^{1/2}.
\end{eqnarray}
The corresponding spectrum for the assumed value of $V/t = 0.01$
is indistinguishable from the spectrum shown in Fig.2(a) for
$V=0$. The differences can be seen on a smaller energy scale, as
in the parts (d) and (e) of Fig.2. When comparing Figs 2(b) and
2(d), one can notice a larger energy gap for $V/t = 0.01$. This is
more clearly seen when comparing the parts (c) and (e). First, for
the assumed value of $V$ the gap is wider than for $V=0$. Second,
the top and bottom band edges become split and shifted away from
the Dirac points. As will be described later, the applied voltage
between the atomic sheets first closes the gap and then opens a
new one with the width increasing with $V$.

\subsection{Effective Hamiltonian}

When only electronic states near the Fermi level (near the Dirac
points) of a neutral graphene are relevant, one can make use of
some effective Hamiltonians to describe the corresponding
electronic spectrum. Such a Hamiltonian can be derived using the
${\bf k}\cdot{\bf p}$ approximation. As a result, the effective
Hamiltonian for states near the $K$ point of the bilayer graphene
takes the form~\cite{mccan, prada}:

\begin{eqnarray}
\label{bg} H_{K} = \mathcal{T}_{0}\varotimes H_K^s
\hspace{3cm}\nonumber\\- \frac{\gamma_{1}}{2}(\mathcal{T}_{x}
\varotimes \sigma_{x} \varotimes S_{0}- \mathcal{T}_{y}\varotimes
\sigma_{y} \varotimes S_{0}),
\end{eqnarray}
where
\begin{eqnarray}
\label{hg} H_{K}^s = v (k_{x}\sigma_{x} + k_{y} \sigma_{y})
\varotimes S_{0} \hspace{2cm} \nonumber \\ + \Delta_{so}
\sigma_{z} \varotimes S_{z} + V \sigma_{0} \varotimes S_{0}.
\end{eqnarray}
The first term on the right side corresponds to two decoupled
atomic monolayers, each of them being described by the Kane
Hamiltonian for a single layer graphene, $H_K^s$. In turn, the
second term describes coupling between the monolayers, with
$\mathcal{T}_{\alpha}$ denoting respectively the unit matrix
($\alpha =0$) and Pauli ($\alpha =x,y,z$) matrices associated with
the layer degree of freedom. In turn, $\sigma_{\alpha}$ ($\alpha =
0, x, y, z$) are the unit ($\alpha = 0$) and Pauli ($\alpha = x,
y, z$) matrices in the pseudo-spin (sublattice) space. Relations
between the parameters of the tight binding and Kane models are:
$v = \sqrt{3}ta/2 = \hbar v_{F}$ ($v_F$ is the carrier velocity at
the Fermi level) and $\Delta_{so} = 3 \sqrt{3} t^\prime$.

The eigenvalues
of Hamiltonian (\ref{bg}) for $V = 0$ take the form
\begin{equation}
\label{ebg12} E_{1,2} = \mp\left[k^{2} v^{2} +
\frac{\gamma_{1}^2}{2} + \Delta_{so}^{2} - \frac{\gamma_{1}}{2}
\sqrt{4 k^{2} v^{2} + \gamma_{1}^{2}}\right]^{1/2}
\end{equation}
and
\begin{equation}
\label{ebg34} E_{3,4} = \mp \left[k^{2} v^{2} +
\frac{\gamma_{1}^2}{2} + \Delta^{2}_{so} + \frac{\gamma_{1}}{2}
\sqrt{4 k^{2} v^{2} + \gamma_{1}^{2}}\right]^{1/2}.
\end{equation}
Electronic spectrum near the point K, described by the above
formula, is shown in Fig.2(b,c), where it is compared with the
spectrum obtained from the full tight binding Hamiltonian. Close
to the K point (gap), spectra from both models coincide very well.

When $V \neq 0$, one finds
\begin{eqnarray}
E_{1,2} = \mp \left[v^{2} k^{2} + \Delta_{so}^{2} + V^{2} + \frac{\gamma_{1}^{2}}{2} \right. \hspace{2.7cm} \nonumber\\
-\left.\frac{1}{2}[(\gamma_{1}^{2} - 4V\Delta_{so})^{2} + 4 v^{2}
k^{2} (\gamma_{1}^{2} + 4 V^{2})]^{1/2} \right]^{1/2},
\end{eqnarray}
\begin{eqnarray}
E_{1^\prime,2^\prime} = \mp \left[v^{2} k^{2} + \Delta_{so}^{2} + V^{2} + \frac{\gamma_{1}^{2}}{2} \right. \hspace{2.7cm} \nonumber\\
-\left.\frac{1}{2}[(\gamma_{1}^{2} + 4V\Delta_{so})^{2} + 4 v^{2}
k^{2} (\gamma_{1}^{2} + 4 V^{2})]^{1/2} \right]^{1/2},
\end{eqnarray}
\begin{eqnarray}
E_{3,4} = \mp \left[v^{2} k^{2} + \Delta_{so}^{2} + V^{2} + \frac{\gamma_{1}^{2}}{2} \right. \hspace{2.7cm} \nonumber\\
+\left.\frac{1}{2}[(\gamma_{1}^{2} - 4V\Delta_{so})^{2} + 4 v^{2}
k^{2} (\gamma_{1}^{2} + 4 V^{2})]^{1/2} \right]^{1/2},
\end{eqnarray}
\begin{eqnarray}
E_{3^\prime,4^\prime} = \mp \left[v^{2} k^{2} + \Delta_{so}^{2} + V^{2} + \frac{\gamma_{1}^{2}}{2} \right. \hspace{2.7cm} \nonumber\\
+\left.\frac{1}{2}[(\gamma_{1}^{2} + 4V\Delta_{so})^{2} + 4 v^{2}
k^{2} (\gamma_{1}^{2} + 4 V^{2})]^{1/2} \right]^{1/2}.
\end{eqnarray}
The above spectrum is shown in Fig.3(d,e), where it is compared
with the corresponding spectrum obtained in the tight binding
model. As before, spectra from tight-binding and effective models
coincide near the K point. Note, the band splitting due to $V$ is
well resolved only in part (e).

Separation of the bands $E_3$ and $E_4$ in the effective model
described above, as well as in the tight binding model, is much
larger than the separation of the bands $E_1$ and $E_2$, see
Fig.2. Therefore, when the electronic states close to the band
edges are relevant and sufficient to describe transport properties
(eg. when the Fermi level is in the gap), one may restrict
considerations to the bands $E_1$ and $E_2$. This leads to a
further simplification of the effective Hamiltonian, as described
in more details in section 4.

\section{Spin Hall and spin Nernst effects}   

Spin Hall and spin Nernst effects correspond to transversal spin
currents induced by electric field and temperature gradient,
respectively. By analogy to the usual Hall and Nernst effects one
may write the density of spin current due to electric field ${\bf
E}$ and temperature gradient ${\bf \nabla}T$ as
\begin{equation}
J_{i}^{s_{n}} = \sum_{j}\left[ \sigma^{s_{n}}_{ij} E_{j} +
\alpha^{s_{n}}_{ij}(-\partial_{j} T)\right]
\end{equation}
where $\sigma^{s_{n}}_{ij}$ (for $i,j=x,y$) is the spin Hall
conductivity with $s_n=(\hbar /2)\sigma_n$ being the $n$-th
component ($n=x,y,z$) of electron spin, while
$\alpha^{s_{n}}_{ij}$ denotes the thermoelectric spin Nernst
conductivity. The two conductivities are not independent and obey
some general relations. Our objective is to find first the
zero-temperature spin Hall conductivity, and then to calculate the
low-temperature thermoelectric spin Nernst conductivity from these
relations, as described below.

The quantum-mechanical operator of spin current density may be
defined as
\begin{equation}
\textbf{J}^{s_{n}}=\frac{1}{2}\left[\textbf{v},s_{n}\right]_{+}
\,,
\end{equation}
where $[A,B]_+=AB + BA$ denotes the anticommutator of any two
operators $A$ and $B$, while $v_i=(1/\hbar )(\partial H/\partial
k_i)$ is the velocity operator ($i=x,y$). The latter operator can
be easily found from the corresponding Hamiltonian [Eqs (1) and
(10)]. In the linear response theory, the frequency-dependent spin
Hall conductivity is then given by the formula~\cite{dyrdal},
\begin{equation}
\label{sigom} \sigma^{s_{z}}_{xy}(\omega)=\frac{e \,\hbar
}{2\omega}{\rm
Tr}\int\frac{d\varepsilon}{2\pi}\frac{d^{2}\mathbf{k}}{(2\pi)^{2}}
\left[v_{x},s_{z}\right]_{+}G_{\mathbf{k}}(\varepsilon +
\omega)v_{y}G_{\mathbf{k}}(\varepsilon),
\end{equation}
where $G_{\mathbf{k}}(\varepsilon )$ is the Green function
corresponding to the appropriate Hamiltonian of the system. When
we restrict considerations to the topological contribution to the
spin Hall current in the d.c. limit, this formula gives exactly
the same result as that based on the Berry phase
calculations~\cite{berry} in momentum space.

It has been shown that the Berry phase leads to an additional term
in the general expression for the orbital
magnetization~\cite{xiao2005}. This correction gives rise to some
contributions to the charge and spin currents~\cite{xiao2007,
chuu}, and also allows to write the relationship between intrinsic
spin Nernst conductivity and intrinsic zero-temperature spin Hall
conductivity, which in the low-temperature regime takes the
form~\cite{chuu}
\begin{equation}
\alpha^{s_{z}}_{xy} = \frac{\pi^{2} k_{B}^{2}}{3 e}\, T
\left.\frac{d \sigma^{s_{z}}_{xy}}{d
\varepsilon}\right|_{\varepsilon=\mu},
\end{equation}
where $T$ stands for temperature and $k_B$ denotes the Boltzman
constant. The latter equation is the spin analog of the Mott
relation for charge transport, and will be used to calculate the
low-temperature spin Nernst conductivity from the zero-temperature
spin Hall conductivity. The derivative in Eq.(21) is taken at the
Fermi level $\mu$. The latter  can be tuned by an external gate
voltage.

Thus, we need to calculate the spin Hall conductivity first. The
relevant derivation depends on the model applied to describe the
corresponding electronic spectrum. Below we present derivation of
the conductivity for the effective Hamiltonian, where analytical
results are available. These results will be compared with those
obtained numerically for the tight binding model.

\subsection{The limit of $V=0$} 

Assume first the limit of $V=0$. To find the spin Hall
conductivity we  start from Eq.(\ref{sigom}) and write it in the
form
\begin{eqnarray}
\sigma_{xy}^{s_{z}}(\omega) = \frac{e}{2 \omega}\int \frac{d
\varepsilon}{2 \pi} \int \frac{d^{2} {\bf k}}{(2 \pi)^{2}} D(\varepsilon + \omega,\varepsilon)\nonumber \\
\times {\small \prod_{n = 1}^{4}} [\varepsilon-E_{n}+\omega+\mu +
i \delta\; {\rm sign}(\varepsilon)]^{-2} \nonumber\\ \times
{\small \prod_{m = 1}^{4}} [\varepsilon-E_{m}+\mu+i \delta \; {\rm
sign}(\varepsilon)]^{-2}.
\end{eqnarray}
Here, $D(\varepsilon + \omega,\varepsilon)$ is defined as
\begin{equation}
\label{tr} D(\varepsilon + \omega,\varepsilon) = {\rm
Tr}\{\left[v_{x},s_{z}\right]_{+}g_{\mathbf{k}}(\varepsilon +
\omega)v_{y}g_{\mathbf{k}}(\varepsilon)\},
\end{equation}
where $g_{\mathbf{k}}(\varepsilon)$ denotes the nominator of the
corresponding Green function $G_{\mathbf{k}}(\varepsilon)$. Taking
the first two terms of the expansion of $D(\varepsilon +
\omega,\varepsilon)$ with respect to $\omega$, one finds
\begin{equation}
D(\varepsilon + \omega, \varepsilon) \simeq i \omega
\chi(\varepsilon),
\end{equation}
with
\begin{eqnarray}
\chi (\varepsilon) = 8 v^{2} \Delta_{so} \left\{v^{2} k^{2}
\left[v^{2}k^{2} + 2
(\Delta^{2} - (\varepsilon + \mu)^{2})\right]\right.\nonumber \\
\left.+(\Delta_{so}^{2} - (\varepsilon + \mu)^{2}) (\gamma_{1}^{2}
+ \Delta_{so}^{2} - (\varepsilon + \mu)^{2})
 \right\}^{2}\nonumber\\
\times \left\{(\Delta_{so}^{2}
- (\varepsilon + \mu)^{2}) (\gamma_{1}^{2} + \Delta_{so}^{2} - (\varepsilon + \mu)^{2})\right.\nonumber\\
+ \left. v^{2} k^{2} \left[v^{2} k^{2} + 2(\gamma_{1}^{2} +
\Delta_{so}^{2} - (\varepsilon + \mu)^{2})\right]\right\}.
\end{eqnarray}
Thus, in the limit of $\omega \rightarrow 0$ one finds the
following expression for the spin Hall conductivity
\begin{equation}
\label{sig} \sigma^{s_{z}}_{xy} = i \frac{e}{2} \int\frac{d
\varepsilon}{2 \pi} \int \frac{d^{2} {\bf k}}{(2 \pi)^{2}}
\mathcal{F}(\varepsilon),
\end{equation}
where
\begin{equation}
\mathcal{F}(\varepsilon) = \frac{\chi
(\varepsilon)}{\prod^{4}_{n=1}[(\varepsilon - E_{n} + \mu + i
\delta\; {\rm sign}(\varepsilon)]^{4}}.
\end{equation}

\begin{figure}[t]
  \includegraphics[width=0.95\columnwidth]{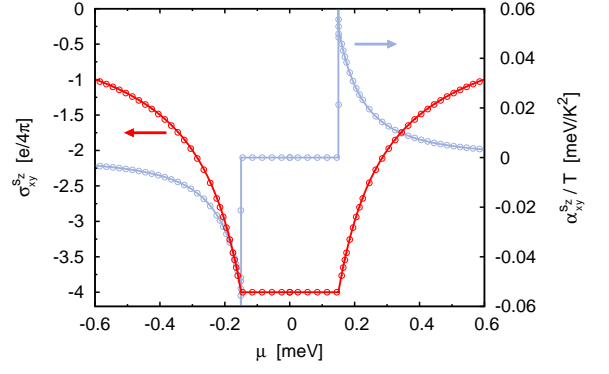}
  \caption{(color on-line) Spin Hall and spin Nernst conductivities of a bilayer graphene
  for $V=0$ and other parameters as in Fig.2. Contributions from
  both $K$ and $K^\prime$ points  are included. The solid lines correspond to the results obtained from
  the effective  Hamiltonian (10), while the dotted ones from the tight binding  model.   }
\end{figure}

Integrating now over $\varepsilon$ one finds
\begin{equation}
\int d \varepsilon \mathcal{F}(\varepsilon) = 2 \pi i \sum_{n}
R_{n} f(E_{n}),
\end{equation}
where $R_n$ ($n=1-4$) are the residua associated with the
corresponding selfenergies (electron bands), and $f(E)$ is the
Fermi distribution function (here for zero temperature). These
residua are equal:
\begin{equation}
R_{1,2} = \pm \frac{4\sqrt{2}v^{2} \Delta_{so} \left[2 v^{2} k^{2}
(\gamma_{1} + \xi)+ \gamma_{1}(\gamma_{1}^{2} - \gamma_{1} \xi + 2
\Delta_{so}^{2})\right]}{\xi^{3} \left(2 v^{2} k^{2} +
\gamma_{1}^{2} - \gamma_{1} \xi + 2 \Delta_{so}^{2}\right)^{3/2}}
\end{equation}
and
\begin{equation}
R_{3,4} = \pm \frac{8 \sqrt{2} v^{2} \Delta_{so} L}{\xi^{3}
\left(2 v^{2} k^{2} + \gamma_{1}^{2} + \gamma_{1} \xi + 2
\Delta_{so}^{2}\right)^{5/2}}
\end{equation}
with
\begin{eqnarray}
L =
- v^{2} k^{2} (3 \gamma_{1}^{3} + \gamma_{1}^{2}\xi + 4 \gamma_{1} \Delta_{so}^{2} - 2 \xi \Delta_{so}^{2}) \nonumber\\
+2 v^{4} k^{4} (\gamma_{1} + \xi)\nonumber\\
- \gamma_{1} (\gamma_{1}^{4} + \gamma_{1}^{3} \xi + 2
\gamma_{1}^{2} \Delta_{so}^{2} + 2 \gamma_{1} \xi \Delta_{so}^{2}
+ 2 \Delta_{so}^{4}) ,
\end{eqnarray}
where $\xi = \sqrt{4 v^{2} k^{2} + \gamma_{1}^{2}}$. Thus, the
conductivity may be written in the form
\begin{equation}
\sigma_{xy}^{s_{z}} = - \frac{e}{4 \pi} \sum_{n}  \int k R_{n}
f(E_{n}) dk .
\end{equation}
Taking into account the integrals
\begin{equation}
\int k R_{1} dk = - \frac{\sqrt{2} (\gamma_{1} + \xi)
\Delta_{so}}{\xi (2 v^{2} k^{2} + \gamma_{1}^{2} - \gamma_{1} \xi
+ 2 \Delta_{so}^{2} )^{1/2}},
\end{equation}
\begin{equation}
\int k R_{3} dk = \frac{\sqrt{2} (\gamma_{1} - \xi)
\Delta_{so}}{\xi (2 v^{2} k^{2} + \gamma_{1}^{2} + \gamma_{1} \xi
+ 2 \Delta_{so}^{2} )^{1/2}},
\end{equation}
and then assuming the appropriate limits of the integration, one
finds the final expressions for the spin Hall conductivity as
presented below.

When the chemical level is inside the gap, $ |\mu| < \Delta_{so}$,
the spin Hall conductivity is equal to
\begin{equation}
\sigma_{xy}^{s_{z}} = - 2 \frac{e}{4 \pi}.
\end{equation}
When   $\Delta_{so} < |\mu| < \sqrt{\gamma_{1}^{2} +
\Delta^{2}_{so}}$,
\begin{equation}
\sigma_{xy}^{s_{z}} = - \frac{2\left( \gamma_{1} + \sqrt{\mu^{2} -
\Delta^{2}_{so}}\right)}{2 \sqrt{\mu^{2} - \Delta^{2}_{so}
}+\gamma_{1}} \frac{\Delta_{so}}{|\mu|} \frac{e}{4 \pi},
\end{equation}
while for $ |\mu| > \sqrt{\gamma_{1}^{2} + \Delta^{2}_{so}}$ one
finds
\begin{eqnarray}
\sigma_{xy}^{s_{z}} = - \frac{2 (\mu^{2} - \Delta_{so}^{2}) -
\gamma_{1}^{2}}{4 (\mu^{2} - \Delta_{so}^{2}) - \gamma_{1}^{2}}
\frac{4 \Delta_{so}}{|\mu|} \frac{e}{4 \pi}.
\end{eqnarray}

The spin Hall conductivity inside the gap is now twice as large as
that in the case of a single-layer graphene.  The general behavior
of the conductivity with position of the Fermi level, shown in
Fig.3 by the solid dark (solid red) line,  is qualitatively
similar to that for a single-layer graphene, i.e., outside the gap
the spin Hall conductivity tends to zero with increasing  $|\mu|$,
while it remains constant and quantized inside the gap. This
behavior is reasonable as the spin Hall conductivity is due to
spin-orbit coupling, which is the same for both atomic monolayers.
Note, the formula derived correspond to one Dirac point, while the
figures include contributions from both Dirac points.

The corresponding spin Nernst conductivity is given by the
following formulas:\\
\\
When $\Delta_{so} < |\mu| < \sqrt{\gamma_{1}^{2} +
\Delta^{2}_{so}}$,
\begin{eqnarray}
\label{11} \alpha^{s_{z}}_{xy} = \pm \frac{\pi}{6} k_{B}^{2} \frac{\Delta_{so}}{\mu^{2}}T \hspace{4.5cm} \nonumber\\
\times \frac{- \gamma_{1}^{2} - 2 \mu^{2} - \frac{4 \gamma_{1}
\mu^{2}}{\sqrt{\mu^{2} - \Delta_{so}^{2}}} + \Delta_{so}^{2}
\left(2 + \frac{3 \gamma_{1}}{\sqrt{\mu^{2} -
\Delta_{so}^{2}}}\right)}{4 \gamma_{1} \sqrt{\mu^{2} -
\Delta_{so}^{2}} + 4 \mu^{2} + \gamma_{1}^{2} - 4
\Delta_{so}^{2}}.
\end{eqnarray}
\\
 For $ |\mu| > \sqrt{\gamma_{1}^{2} + \Delta^{2}_{so}}$,
\begin{eqnarray}
\alpha^{s_{z}}_{xy} = \mp \frac{\pi}{3} \Delta_{so} k_{B}^{2}T
\hspace{4.5cm} \nonumber\\ \times \frac{ 8 \Delta_{so}^{4}+6
\Delta_{so}^{2} \gamma_{1}^{2}+\gamma_{1}^{4}-2 (8
\Delta_{so}^{2}+5\gamma_{1}^{2}) \mu^{2}+8 \mu^{4} }{\mu^{2}
\left( 4 \Delta_{so}^{2}+\gamma_{1}^{2} - 4 \mu^{2}\right)^{2}} .
\end{eqnarray}
\\
In turn, when $\mu$ is in the gap, $|\mu| < \Delta_{so}$, the spin
Nernst conductivity vanishes,  $\alpha^{s_{z}}_{xy} = 0$. The
signs $-$ and $+$ in the above formulas correspond to the case
when the chemical potential is negative or positive, respectively.

Variation of the spin Nernst conductivity with the chemical level
is shown in Fig.~3 by the solid gray (solid blue) line. Similarly
as in a single-layer graphene\cite{czhang2009}, the spin Nernst
conductivity vanishes for the Fermi level inside the gap, when the
system is in the insulating phase, and becomes nonzero for the
Fermi level inside the valence or conduction bands, when the
temperature gradient generates a longitudinal charge current.
Note, the spin Nernst conductivity becomes divergent as $\mu$
approaches edges of the energy gap.

Conductivity in the tight binding model can be obtained in a
similar way, although the corresponding formulas are cumbersome
and will not be presented here. Instead of this we present some
numerical results, which in Fig.3 are shown by the dotted lines.
Note, the spin Hall as well as spin Nernst conductivities in the
effective  model coincide very well with the results obtained from
the tight binding model.

\subsection{The case of $V\ne 0$} 

\begin{figure}[t]
  \includegraphics[width=0.85\columnwidth]{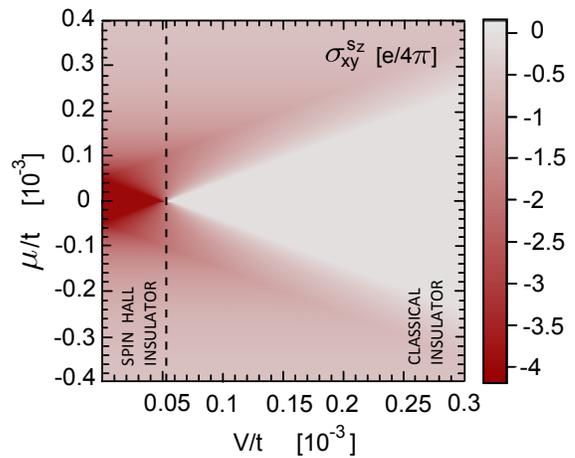}
  \caption{(color on-line) Zero-temperature spin Hall conductivity
  as a function of the vertical voltage $V$ and position of the Fermi
  level for the effective  model described by the parameters given  in Fig.2.
 Contributions from both Dirac points are included. The vertical dashed line corresponds
 to $V$, where transition from the spin Hall insulator to conventional insulator takes place.  }
\end{figure}

Let us consider now the case of $V\ne 0$. The procedure presented
above for the effective model with $V=0$ can be easily extended to
a nonzero vertical bias, $V\ne 0$. The difference is that now the
degeneracy of the bands is lifted and we have 8 different bands,
$n=1 - 8$, which  have to be taken into account. Thus, instead of
Eq.(27) we have now
\begin{equation}
\mathcal{F}(\varepsilon) = \frac{\chi(\varepsilon)
}{\prod^{8}_{n=1}[\varepsilon - E_{n} + \mu + i \delta\, {\rm
sign}(\varepsilon)]^{2}}
\end{equation}
with adequate $\chi(\varepsilon)$. Following the procedure
described above for $V=0$, one can derive the corresponding
analytical formula. These formula, however,  will not be presented
here as they are rather cumbersome, so we present only numerical
results. Moreover,  since the results in the tight binding model
coincide with those obtained with the effective Hamiltonian, as
shown above, we restrict the analysis below to the effective
Hamiltonian.

  \begin{figure}[t]
  \includegraphics[width=0.9\columnwidth]{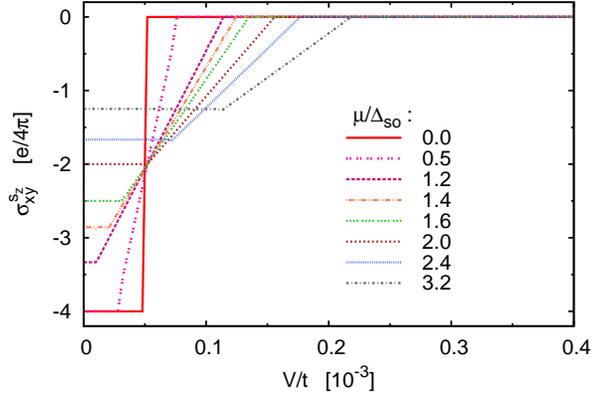}
  \caption{(color on-line) Zero-temperature spin Hall conductivity
  as a function of vertical bias voltage $V$ for indicated values of the
  Fermi
  energy. The curves correspond to crossections of Fig.4 along
  constant $\mu$. The other parameters as in Fig.4. }
\end{figure}

In Fig.4 we show the spin Hall conductivity as a function of the
Fermi level $\mu$ and vertical bias $V$. For $V=0$ we recover the
quantized conductivity in the gap. As $V$ increases, however, the
range of quantized spin Hall conductivity  shrinks and at a
certain value of $V$ (indicated by the dashed line in Fig.4) there
is a transition (at $\mu =0$) from $\sigma_{xy}^{s_{z}}=-4 (e/4
\pi )$ to $\sigma_{xy}^{s_{z}}=0$. This behavior is explicitly
shown in Fig.5, where several cross-sections of Fig.4 along
constant values of $\mu$ are presented. The above transition is
clearly evident for the curve corresponding to $\mu=0$.

  \begin{figure}[t]
  \includegraphics[width=0.8\columnwidth]{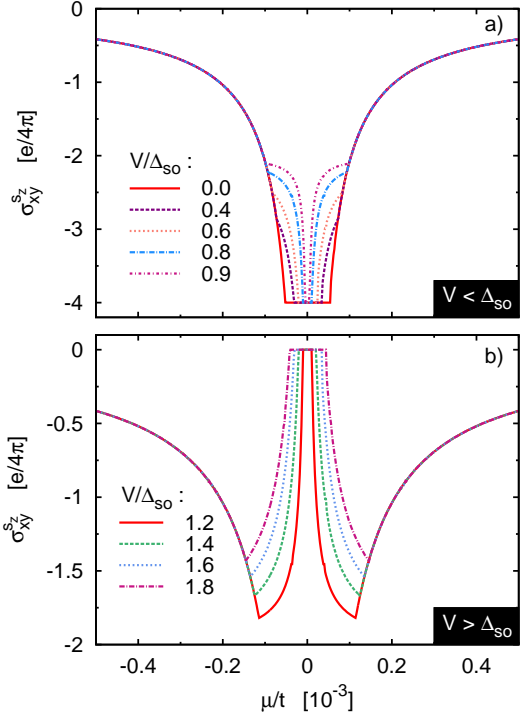}\\
  \caption{(color on-line) Zero-temperature spin Hall conductivity
  as a function of the Fermi level for indicated values of the vertical bias voltage
  $V$. The curves correspond to crossections of Fig.4 along
  constant $V$. The other parameters as in Fig.4.  }
\end{figure}

\begin{figure}[t]
  \includegraphics[width=0.8\columnwidth]{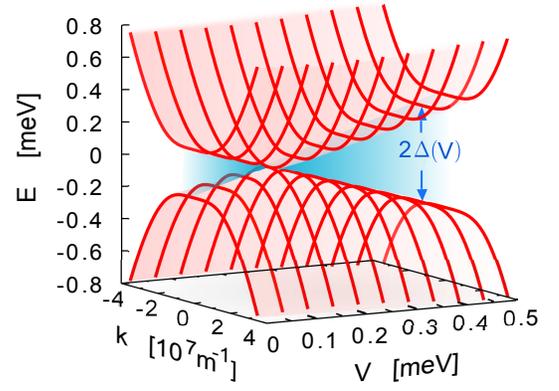}
  \caption{(color on-line) Electronic spectra of the bilayer graphene in the
  vicinity of the Dirac point for different values of $V$. The
  curves show variation of the gap with increasing $V$,  and closure of the gap at some
  critical value of $V$.  The other parameters as in Fig.4}
\end{figure}
  \begin{figure}[h]
  \includegraphics[width=0.75\columnwidth]{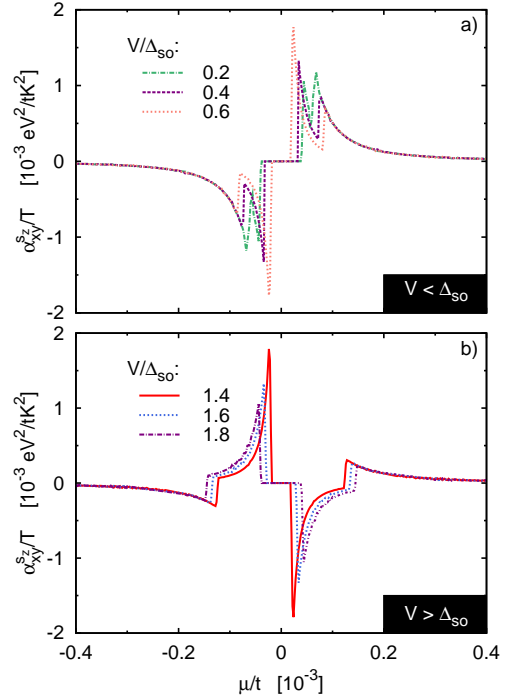}\\
  \caption{(color on-line) Low-temperature spin Nernst  conductivity
  as a function of the Fermi level for indicated values of $V$,
  calculated for the same parameters as in Fig.6.  }
\end{figure}

The transition from topological insulating phase at small voltages
to the normal insulating behavior at large voltages is also
clearly visible in Fig.6, which presents some cross-sections of
Fig.4 along constant values of $V$. This figure shows how the
range of the quantized value of $\sigma_{xy}^{s_{z}}$ changes with
increasing $V$. As $V$ increases starting from $V=0$, width of the
range where $\sigma_{xy}^{s_{z}}$ is quantized shrinks, and at a
certain critical value of $V$ width of this range goes to zero.
The spin Hall conductivity $\sigma_{xy}^{s_{z}}$ at $\mu=0$
changes then from $\sigma_{xy}^{s_{z}}=-4 (e/4 \pi )$ at voltages
smaller then the critical one to $\sigma_{xy}^{s_{z}}=0$ at higher
voltages. This clearly reveals a transition from the topological
insulating phase to the normal insulating behavior (more
information on the topological insulating phases in graphene can
be found eg. in Ref.~[\onlinecite{vozmediano2010}]). From Figs 4
to 6 one could conclude that the gap diminishes with increasing
$V$, becomes totally suppressed at the critical value of $V$, and
then becomes open again at larger voltages. Indeed, this is the
case as shown in Fig.7, where the spectrum near the gap is plotted
for several values of $V$. This figure clearly shows that the gap
becomes closed at the critical value of $V$, and then is open
again at larger values of $V$. The spin Hall conductivity in the
gap above the critical voltage is however suppressed.

The phase transition from the topological spin Hall insulating
phase to the conventional insulator becomes revealed in the spin
Nernst conductivity, too. This is presented in Fig.8, where the
low-temperature spin Nernst conductivity is shown as a function of
the Fermi energy  for the same values of $V$  as in Fig.6. The
range of zero spin Nernst conductivity decreases with increasing
$V$, goes to zero at the critical value of $V$, and then becomes
nonzero again for larger values of $V$.

\section{Low-energy effective Hamiltonian}  

As we have already mentioned above, separation of the bands $E_3$
and $E_4$ (or $E_{3,3^\prime}$ and $E_{4,4^\prime}$) is much
larger than separation of the bands $E_1$ and $E_2$ (or
$E_{1,1^\prime}$ and $E_{2,2^\prime}$). The latter determines the
energy gap induced by the spin-orbit coupling (see Fig.2). When
only the electron states near the Fermi level are relevant and the
Fermi level is in the gap or close to it, one may further reduce
the effective Hamiltonian to include explicitly the bands
$E_{1,1^\prime}$ and $E_{2,2^\prime}$, and the other bands only
via effective parameters of the corresponding reduced effective
model. The relevant reduced low-energy effective Hamiltonian takes
the form~\cite{mccan}:
\begin{eqnarray}
\label{hlow} H^{r}_{K} = \left[\begin{array}{c c c c}
                                 \Delta_{so} + V & 0 & - \frac{\hbar^2k^{2}_{-}}{2 m} & 0 \\
                                 0 & - \Delta_{so} + V & 0 & - \frac{\hbar^2k^{2}_{-}}{2 m} \\
                                 - \frac{\hbar^2k^{2}_{+}}{2 m} & 0 & - \Delta_{so} - V& 0 \\
                                0  & - \frac{\hbar^2k^{2}_{+}}{2 m} & 0 & \Delta_{so} - V\\
                                \end{array}\right],
\end{eqnarray}
where $k_{\pm} = k_{x} \pm ik_{y}$ and $m = \gamma_{1}/2 v_{F}$ is
the effective electron mass.  In the following we consider some
special cases.

\subsection{The case of  $V = 0$}

The corresponding eigenvalues of the Hamiltonian (41) are then
equal to
\begin{equation}
\label{lowEnV0} E_{1,2} = \mp  \left[\Delta_{so}^{2} +
\left(\frac{\hbar^2k^{2}}{2 m }\right)^2\right]^{1/2},
\end{equation}
see the inset in Fig.9. Using the notation introduced in the
preceding section we find
\begin{equation}
\sigma_{xy}^{s_{z}} = i \frac{e}{2} \int \frac{d \varepsilon}{2
\pi} \int \frac{d^{2} {\bf k}}{(2 \pi)^{2}} \frac{\chi
(\varepsilon)}{ \prod_{n=1}^{n=2}[\varepsilon - E_{n} + \mu + i
\delta {\rm sgn} ({\varepsilon})]^{4}},
\end{equation}
where
\begin{eqnarray}
 \chi (\varepsilon) = \frac{4\Delta_{so} \hbar^4k^{2}}{ m^{2}} \left[\Delta_{so}^{2} -
 (\varepsilon + \mu)^{2} + \left(\frac {\hbar^2k^{2}}{2m}\right)^2\right]^{2}.
\end{eqnarray}
Performing the integration over $\varepsilon$ and then over ${\bf
k}$, one arrives at the following analytical formulas for the spin
Hall conductivity:
\begin{equation} \sigma_{xy}^{s_{z}} = - \frac{2
\Delta_{so}}{|\mu|} \frac{e}{4 \pi}
\end{equation}
for $|\mu| > \Delta_{so}$, and
\begin{equation}
\sigma_{xy}^{s_{z}} = - 2\frac{e}{4 \pi}
\end{equation}
when the chemical level is inside the gap, $|\mu| < \Delta_{so}$.

\begin{figure}[t]
  \includegraphics[width=1.0\columnwidth]{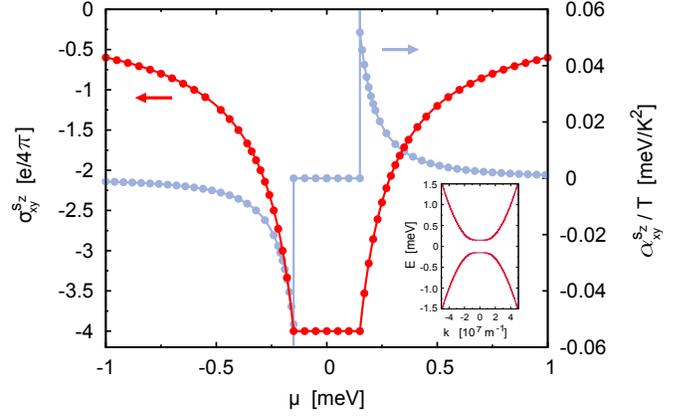}
  \caption{(color on-line) Spin Hall and spin Nernst conductivities of a bilayer graphene
  for the low energy reduced Hamiltonian with the parameters corresponding to those assumed in Fig.2.
  Contributions from
  both $K$ and $K^\prime$ points are included. The solid lines correspond to the results obtained from
  the Hamiltonian (10), while the dotted ones from the reduced Hamiltonian (41). The inset shows the
  energy spectrum  corresponding to Eq.(42). }
\end{figure}

The corresponding low-temperature spin Nernst conductivity is then
given as follows.  For $|\mu|
> \Delta_{so}$ one finds
\begin{equation}
\alpha_{xy}^{s_{z}} = \mp \frac{\pi}{6} k_{B}^{2}
\frac{\Delta_{so}}{\mu^{2}} T,
\end{equation}
while for  $\mu$ inside the gap $\alpha_{xy}^{s_{z}} = 0$

The above results for both spin Hall and spin Nernst
conductivities are shown by the dotted lines in Fig.9, where they
are compared with the corresponding results obtained from the
effective Hamiltonian (10) (solid lines). There is a nice
agreement between the results. We note that the divergency of the
spin Nernst conductivity when the Fermi level approaches the band
edges, observed in the tight-binding and  effective Hamiltonian,
is not reproduced by the reduced Hamiltonian.

\subsection{The case of $V \neq 0$}
When $V\ne 0$  the eigenvalues of the Hamiltonian (41) take
the  form,
\begin{equation}
E_{1,2} = \mp \left[ (V + \Delta_{so})^{2}
+\left(\frac{\hbar^2k^{2}}{2m}\right)^2 \right]^{1/2},
\end{equation}
\begin{equation}
E_{1^\prime,2^\prime} = \mp \left[ (V -
\Delta_{so})^{2}+\left(\frac{\hbar^2k^{2}}{2m}\right)^2\right]^{1/2}.
\end{equation}
The nonzero $V$ leads to splitting of the electron bands, as
already discussed above.

When $|\mu| > (V + \Delta_{so})$, the spin Hall conductivity is
then given by
\begin{equation}
\sigma_{xy}^{s_{z}} = - \frac{e}{4\pi} \frac{2\Delta_{so}}{|\mu|}
\end{equation}
When $|V-\Delta_{so}| < |\mu| < (V + \Delta_{so})$,
\begin{equation}
\sigma_{xy}^{s_{z}} = - \frac{e}{4\pi} \left(1 - \frac{V -
\Delta_{so}}{|\mu|}\right)
\end{equation}
Finally, when $\mu$ is inside the gap,
\begin{eqnarray}
\sigma_{xy}^{s_{z}} = 0 \quad {\rm for} \quad V>\Delta_{so} \\
\sigma_{xy}^{s_{z}} = -2 \frac{e}{4 \pi} \quad {\rm for} \quad
V<\Delta_{so}.
\end{eqnarray}

The corresponding low-temperature spin Nernst conductivity is
given by
\begin{equation}
\alpha_{xy}^{s_{z}} =  \frac{\pi}{6} k_{B}^{2}\, T
\frac{\Delta_{so}}{\mu |\mu|}
\end{equation}
 for $|\mu| > V + \Delta_{so}$, and
 \begin{equation}
 \alpha_{xy}^{s_{z}} = - \frac{\pi}{12} k_{B}^{2}\, T\frac{1}{|\mu|\mu} (V -
 \Delta_{so})
 \end{equation}
 for $|V - \Delta_{so}| < |\mu| < (V + \Delta_{so})$.
In turn, the spin Nernst conductivity vanishes inside the energy
gap.

Behavior of spin Hall and spin Nernst conductivities with position
of the Fermi level and bias voltage $V$ almost coincides with that
presented in Figs 4 to 6 and 8, and therefore will not be present
here. As in the ${\bf k}\cdot{\bf p}$ model, one observes the same
transition between the spin Hall insulator and classical insulator
as $V$ increases. As already mentioned above, there is no
divergency of the spin Nernst conductivity at the band edge for
the low energy effective model considered here.

\section{Summary}

We have calculated analytically as well as numerically the spin
Hall and spin Nernst conductivities in a  bilayer graphene. To
describe the relevant electronic spectrum  we have assumed the
tight binding model as well as some simplified effective
Hamiltonians relevant for states close to the Dirac points. Both
spin Hall and spin Nernst effects consist in transverse spin
accumulation (spin current). However, as the spin Hall effect is
due to external electric field, the Nernst effect is due to a
temperature gradient. Assuming intrinsic spin orbit interaction,
we have found the intrinsic contributions to both effects.
Generally, the spin Hall conductivity in the energy gap of a
bilayer graphene is twice as large as that for a single-layer
graphene. When external voltage is applied between the two atomic
sheets, we found a transition from the spin Hall insulator phase
to the conventional insulator behavior. The energy gap at the
transition point is closed.

We have compared results obtained from the tight-binding model
with those derived from the effective Hamiltonians. From this
comparison we arrived at the conclusion, that both spin Hall and
spin Nernst conductivities are well described by the effective
Hamiltonians, which allow derivation of  some analytical results.

\subsection*{Acknowledgment}

This work has been supported in part by the European Union under European
Social Fund 'Human - best investment' (PO KL 4.1.1) and in part by
funds of the Ministry of Science and Higher Education as a
research project in years 2010-2013 (No. N N202 199239). The
authors acknowledge valuable discussions with V.K. Dugaev.


\end{document}